\documentclass{article}

\usepackage{PRIMEarxiv}

\usepackage[utf8]{inputenc} 
\usepackage[T1]{fontenc}    
\usepackage{hyperref}       
\usepackage{url}            
\usepackage{booktabs}       
\usepackage{amsfonts}       
\usepackage{nicefrac}       
\usepackage{microtype}      
\usepackage{lipsum}
\usepackage{fancyhdr}       
\usepackage{graphicx}       
\usepackage{longtable}
\graphicspath{{media/}}     

\title{Study of the motion of Globular Clusters relative to the plane and center of the 	Galaxy based on Gaia and HST data
\textbf{}}

\author{G.S. Karapetian\textsuperscript{1}, A.P. Mahtessian\textsuperscript{1*}, M. A. Hovhannisyan\textsuperscript{2}, L.A. Mahtessian\textsuperscript{2} and L.E
Byzalov\textsuperscript{3}
 \\
  Affiliation \\
\textsuperscript{1} Byurakan Astrophysical Observatory after V. Ambartsumian NAS of the Republic \\ of  Armenia,  Byurakan, Aragatzotn Province , Republic of Armenia, 0213  \\
\textsuperscript{2 }Institute of Applied Problems of Physics NAS of the Republic of Armenia
25 Hrachya Nersissian Str.,\\ Yerevan, Republic of Armenia, 0014\\
\textsuperscript{3 } University of Waterloo, Ontario, Canada\\  200 University Ave W, Waterloo, ON N2L 3G1\\
  \texttt{* amahtes@gmail.com} \\
}

\usepackage{xcolor}

\begin{document}
\maketitle

\begin{abstract}
The authors of the article set out to investigate the motion of Globular Star Clusters (GCs) of the Milky Way relative to the plane and center of the Galaxy. For this study, we used data from 164 Galactic Globular Clusters derived from Gaia, HST, and literature studies.

The study calculated the average GC motion velocities relative to the Galactic plane in the northern and southern regions depending on the galactic latitude b and distance from the Galactic plane.

The work consisted of using the GC parameters to find the proper motion of clusters along the Galactic latitude b and the velocities V↑ and V↓ of cluster motion in the perpendicular direction relative to the Galactic plane for the northern and southern regions, respectively, and determine the average values of these velocities.

The velocities V↑ and V↓ of GC motion were also found depending on the distance to the Galaxy plane and their average values were determined.

The work also determined the average velocity VR of motion of the GC relative to the center of the galaxy.

The study yielded the following results:

    1. The average velocity of the GC relative to the Galactic plane for the northern      region of the Galaxy                     has a positive value of approximately 30.36 ±         15.29 km/s (N=42) for the range 0°<b<13°.

2. For the southern region, the following result was obtained: the average velocity   of the GC movement relative to the plane in the southern direction is -17.83 ±13.54 km/s (N=50) for the range 0°>b>-13°.

3. The average velocity of GC motion relative to the Galactic plane for the northern region    has a positive value of 27.58 ± 16.93 km/s (N=35) for the range of distances from the plane 0 - 1.4 kpc

4. For the southern region: the average velocity    of the GC movement relative to the plane in the southern direction is -18.08 ±14.89 km/s (N=42) for the range of distances to the plane 0 - 1.4 kpc

5. The average velocity of the GC relative to the Galactic center for the range 0 to 12 kpc is 31.70 ± 14.74 km/s (N=119).

Thus, we can state with some confidence that GCs, on average, are moving away both from the plane and from the center of the Galaxy.

\end{abstract}


\section{Introduction}
The new Gaia Mission EDR3 and ED3 data represent improved parameters of the photometric and astrometric catalogs compared to previous versions. In particular, the inaccuracies in parallaxes $\omega$ and proper motions $\mu$ were reduced on average by a factor of two, and systematic inaccuracies were reduced even more. Already DR2 has made it possible to calculate the average parallaxes (\cite{ChenRicher2018}, \cite{ShaoLi2019} and proper motions of almost all Globular Clusters (\cite{Gaia2018}, \cite{VasilievBaumgardt2021}, \cite{Vasiliev2019} of the Milky Way. Since globular cluster stars have similar kinematic properties and parallaxes, \cite{Baumgardt2021} used these data sets (up to tens of thousands of stars in the richest clusters) to determine whether these stars belong to the GC and calculate their general parameters.

Thus, a catalog of 164 Globular Clusters (GCs) with the necessary parameters was created, which we used to study the velocities of these GCs relative to the plane and center of the Milky Way Galaxy.

First, we calculated the velocities of movement of all GCs in the orthogonal direction relative to the plane for the northern and southern regions of the Galaxy. Further, in Section 2.1 we calculated the average GC movement velocities depending on the Galactic latitude b, and in Section 2.2, separately, on the distance of the GC to the Galactic plane. Figures 1, 2 show the dependence of GC velocities on galactic latitude for the northern and southern regions of the Galaxy. Tables 1 and 2 summarize the main findings of the study.

\section{Method}
\label{sec:headings}

As the main sample of Globular star clusters in the Milky Way Galaxy, we used the catalog of 164 GCs given in the article \cite{Baumgardt2021} in the Data Availability section (\url{https://people.smp.uq.edu.au/HolgerBaumgardt/globular/orbits.html}. 

The table gives the orbital parameters of 164 Galactic Globular Clusters as derived from the Gaia EDR3 proper motions and radial velocities. The Table contains all confirmed Milky Way Globular clusters with proper motion and radial velocity information.\\ 
The majority of the 164 GCs are located in the central part of the Galaxy, so we selected GCs from the table according to the following parameters: GCs with galactic longitude l in the range 0°<l<90° and 270°<l<360°. Thus, out of 164, 151 GCs remain in the table. Next, we selected a GC with a galactic latitude of 13°>b>-13° (92 GC) and separately, a distance to the Galactic plane from 1.4 kpc to -1.4 kpc (77 GC).

\subsection{Dependence of GC velocities V↑ and V↓ on galactic latitude b. }
The sample for 13°>b>-13° consists of 92 GCs (Table 1).
In the original table, \cite{Baumgardt2021} the following GC parameters are given:

\begin{itemize}
    \item RA: Right ascension
    \item DEC: Declination
 \item l: Galactic longitude
    \item b: Galactic latitude
    \item ${R_{\odot}}$: Distance from Sun
      \item RGC: Distance from the Galactic center
       \item RV: Radial velocity
       \item $\mu\alpha\cos\delta: $Proper motion in right ascension
       \item $\mu\delta$: Proper motion in declination
       \item $\rho\mu\alpha\mu\delta$: Correlation coefficient
      \item X: Distance from the Gal. center in direction of Sun
       \item Y: Distance from the Gal. center in direction of Solar motion
       \item Z: Distance above/below galactic plane
       \item U: Velocity in X direction
       \item V: Velocity in Y direction
        \item W: Velocity in Z direction
       \item RPeri: Average perigalactic distance
       \item RApo: Average apogalactic distan
\end{itemize}

As stated in the article \cite{Baumgardt2021} distances ${R_{\odot}}$ to globular clusters are determined either by using eclipsing binaries (\cite{Kaluzny2013}, \cite{Thompson2020} or through fits of their color-magnitude diagrams (CMDs) with theoretical isochrones e.g. \cite{Ferraro2000}, \cite{Dotter2010}, \cite{Gontcharov2019}, \cite{Valcin2020}, by using variable stars that follow known relations between their periods and absolute luminosities like RR Lyrae stars e.g. \cite{Bono2020}, \cite{Hernitschek2019}, Type II Cepheid \cite{Matsunaga2006} or Mira type variables \cite{Feast2002}. Finally, it is possible to determine distances by comparing the magnitudes of main-sequence stars with stars of similar metallicity in the solar neighborhood, the so-called subdwarf method e.g. \cite{Reid1998}, \cite{Cohen2011} or kinematically by comparing line-of-sight and proper motion velocity dispersion profiles in globular clusters e.g. \cite{McNamara2004}, \cite{Ven2006}, \cite{Watkins2016}. The latter method has the advantage that the derived distance is not influenced by the reddening of the cluster.
For GCs with accurately measured radial velocities and dispersion of proper motions, the distance to the GC can be determined by achieving a better fit with a theoretical cluster model e.g. \cite{Hénault2019}.

RV radial velocities were calculated by \cite{Baumgardt2018} by combining literature data with ESO and Keck data archives. \cite{Baumgardt2019} also added radial velocities from Gaia DR2 and the Anglo-Australian Observatory (AAO) to the table. The radial velocities given in \cite{Kamann2018} were also added based on MUSE data.

Own motions were added to the table from Gaia EDR3 or HST.

For this work, we used the following table parameters: RA, DEC, l, b, RGC, $\mu$$\alpha$cos$\delta$, $\mu$$\delta$, Z and W. The parameters l, b, RGC, Z and W in the table are given with rounded precision, so these parameters were calculated by us using the appropriate formulas. In particular, l and b are given in the table with an accuracy of 3 decimal digits after comma, and $\alpha$ and $\delta$ are given with an accuracy of 5 decimal digits after comma, so we considered it appropriate to obtain l and b ourselves by converting coordinates using the formulas:

\begin{equation}
\sin(b)=sin(\delta_{NGP})*sin(\delta)+cos(\delta_{NGP})*cos(\delta)*cos(\alpha-\alpha_{NGP})
\end{equation}
\begin{equation} 
\tan(l_{NCP}-l)=\frac {cos(\delta)*sin(\alpha-\alpha_{NGP})}{sin(\delta)*cos(\delta_{NGP})-cos(\delta)*sin(\delta_{NGP})*cos(\alpha-\alpha_{NGP})}
\end{equation}
where:
\begin{itemize}
    \item $\alpha$ and $\delta$ – equatorial coordinates of the GC,
    \item $\delta_{NGP}$ and $\alpha_{NGP}$ are the equatorial coordinates of the north galactic pole,
    \item $l_{NCP}$ – Galactic longitude of the North celestial pole,
    \item l and b are the galactic coordinates of the GC.
\end{itemize}
The proper motion $\mu$l, $\mu$b were calculated based on $\mu\alpha$,
$\mu\delta$ as follows: first, l and b are calculated based on the above formulas and then l'
and b' (GC galactic coordinates in a year) using $\alpha$'=$\alpha$+$\mu\alpha$ and $\delta$'=$\delta$+$\mu\delta$. Next we get $\mu$l= (l'- l), $\mu$b= b'- b.

Using $\mu$b, we obtain the velocity of movement V↑ of the GC in the northern region relative to the Galaxy plane according to the formula:

    V↑ = $(R_{\odot}+RV)*sin(b+{\mu}b) - R_{\odot}*sin(b)$

where:
 \begin{itemize}
 \item ${R_{\odot}}$ - Distance from the Sun
 \item RV - Radial velocity of GC
\end{itemize}
\color{black}
Next, we convert these velocities to km/s.
Thus, we obtained the velocities V↑ of the GC motion in the direction orthogonal to the galactic plane. We, then  sorted them in Table 1 in descending order of b.
 For the southern region of the Galaxy, we will do the same with negative values of the indicated ranges b and denote the velocities V↓.
Figures 1 and 2 show the GC velocities V↑ and V↓ in the northern and southern regions of the Galaxy.

Thus, it can be argued that the GCs of the northern region in the range 0°<b<13° on average move away from the Galactic plane. The average velocity of the GC's receding from the Galactic plane in the northern region is 30.36 ± 15.29 km/s (N=42). Similarly, GCs in the southern region of the Galaxy in the range -13°<b<0° on average move away from the Galactic plane at a velocity of 17.85 ± 13.76 km/s (N=50). Application of Student's t-test shows that the GC velocities V↑ and V↓ of the northern and southern regions of the Galaxy in the ranges 0°<b<13° and -13°<b<0° differ at the level of 0.0205.
\subsection{GC velocities V↑ and V↓ relative to the Galaxy plane depending on the distance to the plane. }
In the previous section, we obtained the GC movement velocities depending on the galactic latitude b.
Now let's move on to determining the GC velocities relative to the distance to the Galactic plane (Dpl). The GC sample from the general table for the range of distances to the plane from 0 to 1.4 kpc contains 77 GCs (Table 2).
GC velocities V↑ and V↓ for the northern and southern regions of the Galaxy are shown in Fig. 3 and 4.

On average, the GCs of the northern and southern regions of the Galaxy move away from the Galactic plane. The average velocity of the GCs receding from the plane for the northern region of the Galaxy is 27.74 ± 17.13 km/s (N=35) and for the southern region  -17.41 ± 14.51 km/s (N=42). Application of Student's t-test shows that the GC velocities V↑ and V↓ of the northern and southern regions of the Galaxy for the distance ranges from the plane 0 < Dpl < 1.4 kpc and -1.4 < Dpl <0 kpc differ at the level of 0.0386.
The dependence of the GC velocities V↑ and V↓ over the entire range -1.4 < Dpl < 1.4 kpc is shown in Fig. 5.

As can be seen from the figure 5, there is an obvious dependence between the parameters discussed. Let us evaluate the significance of the correlation. To do this, we use the value
\begin{equation}
t= R\sqrt{\frac{n-2}{1-R^2}}
\end {equation}
which obeys the Student distribution.
Here,
\begin{equation}
R(V,Dpl)=\frac{\sum(V_i-<V>)(D_{pli}-<D_{pl}>)}{(n-1)\sigma _{V}\sigma_{Dpl}}
\end{equation}

- assessment of the correlation between the values of V and Dpl, $\sigma_V$ and $\sigma_{Dpl}$ - standard deviations of the corresponding values.
\begin{equation} 
\sigma_V=\sqrt\frac{\sum_{i=0}^{n}{(V_i-<V>)^2}}{n-1},  \,\,\,\,\,
\sigma_{Dpl}=\sqrt\frac{\sum_{i=0}^{n}{(Dpl_i-<Dpl>)^2}}{n-1},
\end{equation}

We get t=2.55, from which it follows that the significance of the correlation is high $\alpha$ =1-P $\approx$ 0.01.
Thus, we can conclude that GCs far from the Galactic plane, on average, have relatively high speeds of receeding  from the Galactic plane.
\subsection{GC motion relative to the Galactic center. }
The result obtained in the previous sections that GCs on average move away from the Galactic plane suggests that it is possible that GCs on average can also move away from the Galactic center, so this work attempts to test this assumption.
 Using the table data from 151 GCs, the distances of the GC from the center of the Galaxy and the velocity of movement of the GC (VR) relative to the center of the Galaxy were calculated.
To do this, we calculated the distance of each GC from the center of the Galaxy for two points of location of these GCs: the “current” (or rather at the moment of observation) distance of the GC from the center R and the distance calculated after a 1 year R'. The difference between these distances gives us the velocity of movement of the GC relative to the center, i.e. VR = R' - R. In the table, the current distance of the GC from the center of the Galaxy RGC is given rounded to the second decimal number, so we considered it appropriate to calculate these values ourselves.
Using simple trigonometry, the velocities of globular clusters relative to the center of the Galaxy are calculated.
The current distance of the GC from the Galactic center is calculated using the following formula:
\begin{equation}
R=\sqrt{(D_{CL}*sin(b))^2+(D_{GC}*sin(l))^2+(D_{CL}*cos(b)-D_{GC}*cos(l))^2}
\end{equation}
 where: 
\begin{itemize}
    \item $D_{GC}$(kpc)– distance from the Sun to the center of the Galaxy;
    \item b – galactic latitude of the Globular Cluster (GC);
    \item l – galactic longitude of the Globular Cluster (GC);
    \item $D_{Cl}$ – distance from the Sun to the center of the GC;
    \item R - distance from the center of the Galaxy to the center of the GC;
\end{itemize}
The distance of the GC from the Galactic center in a 1 year is calculated using the following formula: 
\begin{equation}
R'=\sqrt{(D'_{CL}*sin(b'))^2+(D_{GC}*sin(l'))^2+(D'_{CL}*cos(b')-D_{GC}*cos(l'))^2}
\end{equation}
where:
\begin{itemize}
    \item $D'_{Cl}=D_{Cl}+RV$  distance from the Sun to the GC in a year;
    \item b' = b+ $\mu$b - galactic latitude in a year;
    \item l' = l+ $\mu$l - galactic longitude in a year;
    \item RV – radial velocity of the GC relative to the Sun.
\end{itemize}
The average velocity of the GC relative to the Galactic center for the range 0 to 12 kpc is 31.70 ± 14.74 km/s (N=119).
Based on the calculations obtained, it can be assumed that GCs, on average, are also moving away from the Galactic center. 
\section{Conclusions and discussion}
Thus, within the framework of this work, the velocities of movement V↑ and V↓ of the globular clusters of the northern and southern regions of the Galaxy were calculated depending on the Galactic latitude b and the distance to the Galactic plane. The GC motion velocities relative to the Galactic center were also calculated.
The following results were obtained:

• GCs of the northern region in the range 0°<b<13° on average move away from the Galactic plane. The average velocity of the GC's receding from the Galactic plane in the northern region is 30.36 ± 15.29 km/s (N=42).

• GCs in the southern region of the Galaxy in the range 0>b>-13° on average move away from the Galactic plane at a velocity of 17.85 ± 13.76 km/s (N=50).

• GCs in the northern region of the Galaxy in the range of distances to the plane from 0 to 1.4 kpc on average move away from the plane of the Galaxy. The average velocity of receding of the GC in the northern region from the Galactic plane is 27.74 ± 17.13 km/s (N=35).

• GCs in the southern region of the Galaxy in the range of distances to the plane from 0 to 1.4 kpc on average move away from the plane of the Galaxy. The average velocity of receding of the GC in the southern region from the Galactic plane is 17.41 ± 14.51 km/s (N=42).

• The average velocity of receding of the GC from the Galactic center for the range 0 to 12 kpc is 31.70 ± 14.7 km/s. (N=119). 

 Thus, it can be stated that the population of the second type of Galaxy, using the example of Globular clusters, at least in the central part, is expanding.
Refinement of the results obtained and further research will be facilitated by the presence of a larger number of studied clusters with more accurate parameters.
\section {Data availability}
For the research in this paper, we used the table of 164 well-studied Globular Clusters of the Milky Way Galaxy given by \cite{Baumgardt2021}.
The table
gives the orbital parameters of Galactic Globular Clusters as derived from the Gaia proper motions and radial velocities. The Table contains all confirmed Milky Way globular clusters with proper motion and radial velocity information.
The distances and the other parameters of Galactic globular clusters can be obtained from the following webpage:
https://people.smp.uq.edu.au/HolgerBaumgardt/globular/orbits.html

\bibliographystyle{unsrt}  
\bibliography{references}  

\newpage
\begin{figure}[hbt!]
    \centering
    \includegraphics[width=1\linewidth]{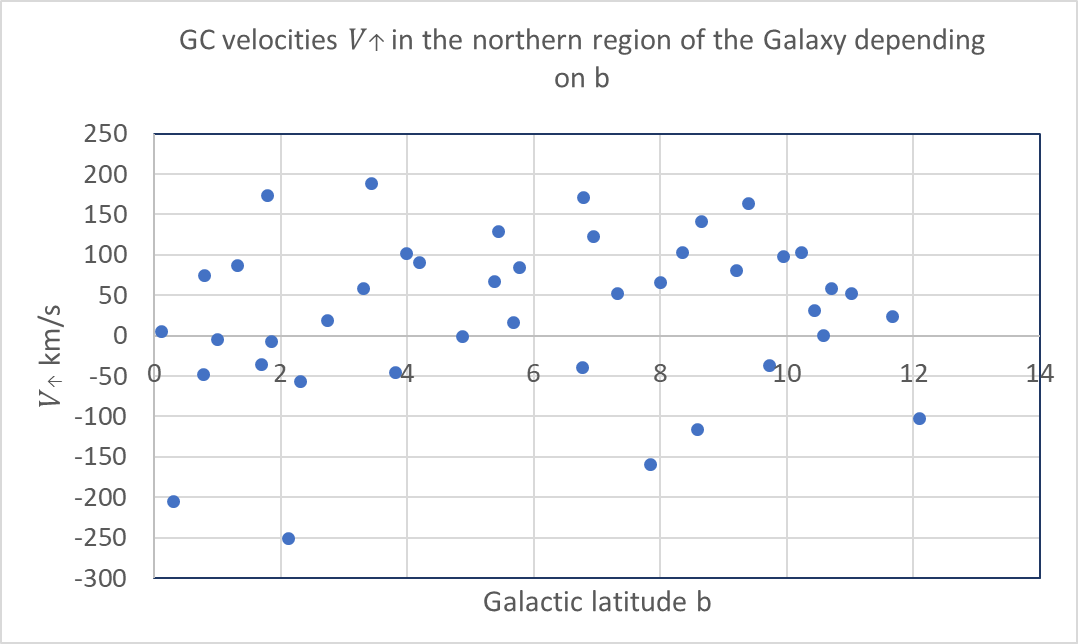}
    \caption{GC velocities V↑ in the northern region of the Galaxy depending on coordinate b.
As can be seen from the figure, most GCs have a positive velocity, i.e. moving away from the galactic plane.
}
    \label{fig:1}
\end{figure}
\begin{figure}
    \centering
    \includegraphics[width=1\linewidth]{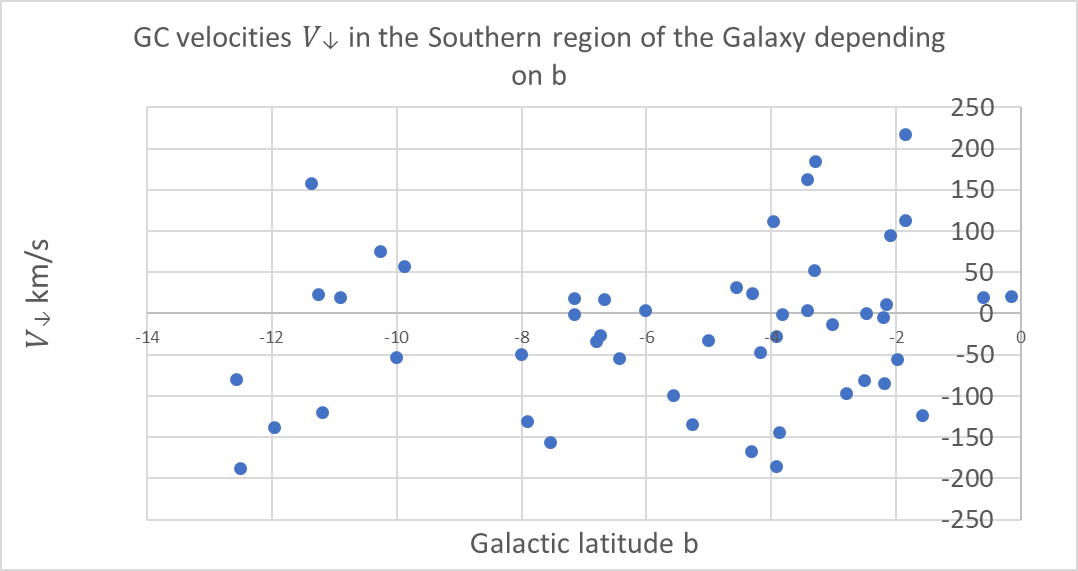}
    \caption{ GC velocities V↓ in the southern region of the Galaxy depending on coordinate b. Most GCs have a negative velocity, i.e. moving away from the galactic plane.}
    \label{fig:2}
\end{figure}
\begin{figure}
    \centering
    \includegraphics[width=1\linewidth]{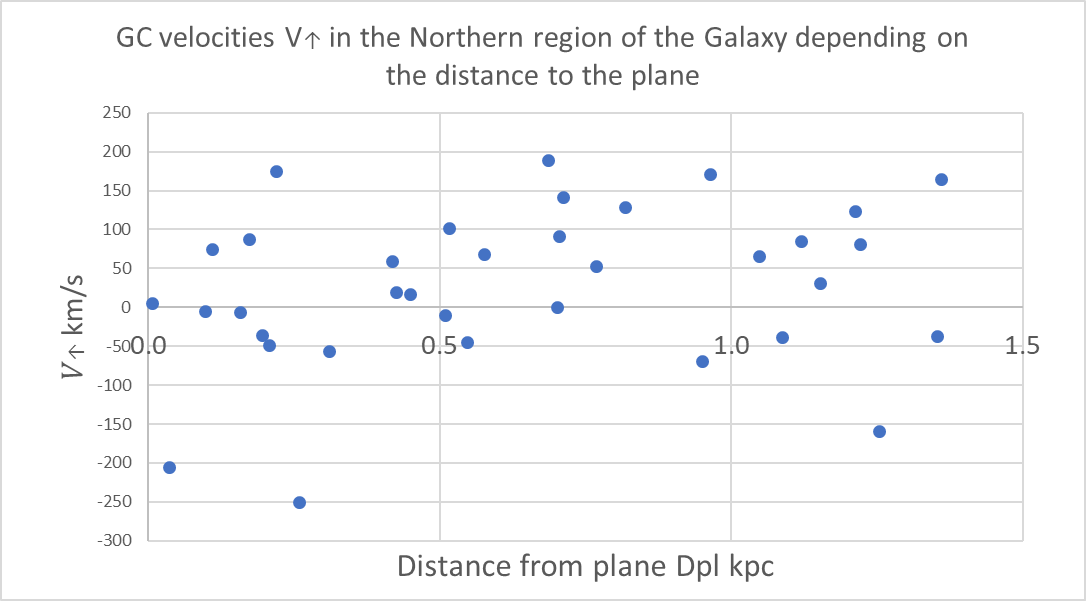}
    \caption{ GC velocities V↑ in the northern region of the Galaxy depending on the distance to the Galactic plane. As can be seen from the figure, most GCs have a positive velocity, i.e. moving away from the galactic plane.}
    \label{fig:3}
\end{figure}
\begin{figure}
    \centering
    \includegraphics[width=1\linewidth]{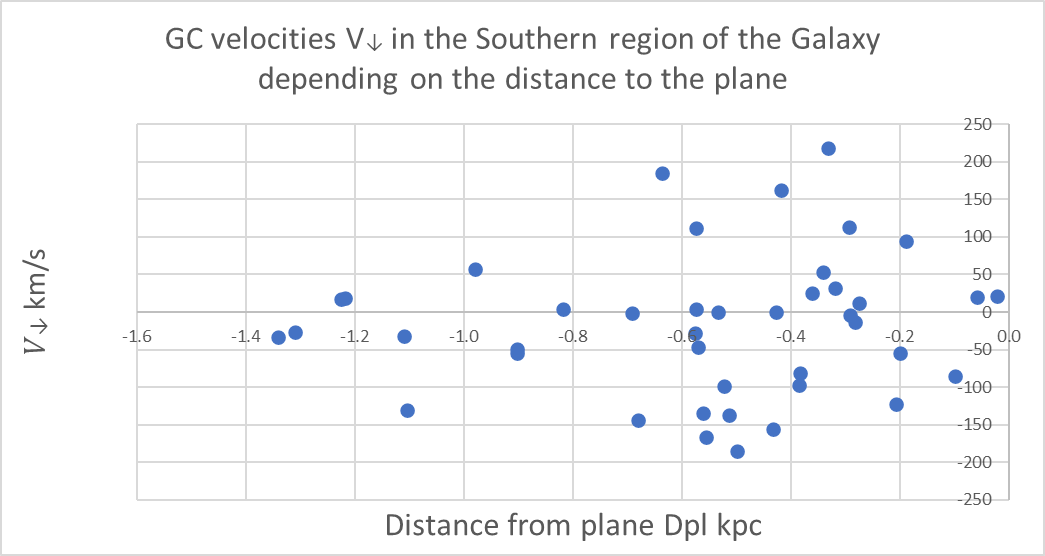}
    \caption{ GC velocities V↓ in the southern region of the Galaxy depending on the distance to the Galactic plane Dpl. Most GCs have a negative velocity, i.e. move away from the plane.}
    \label{fig:4}
\end{figure}
\begin{figure}
    \centering
    \includegraphics[width=1\linewidth]{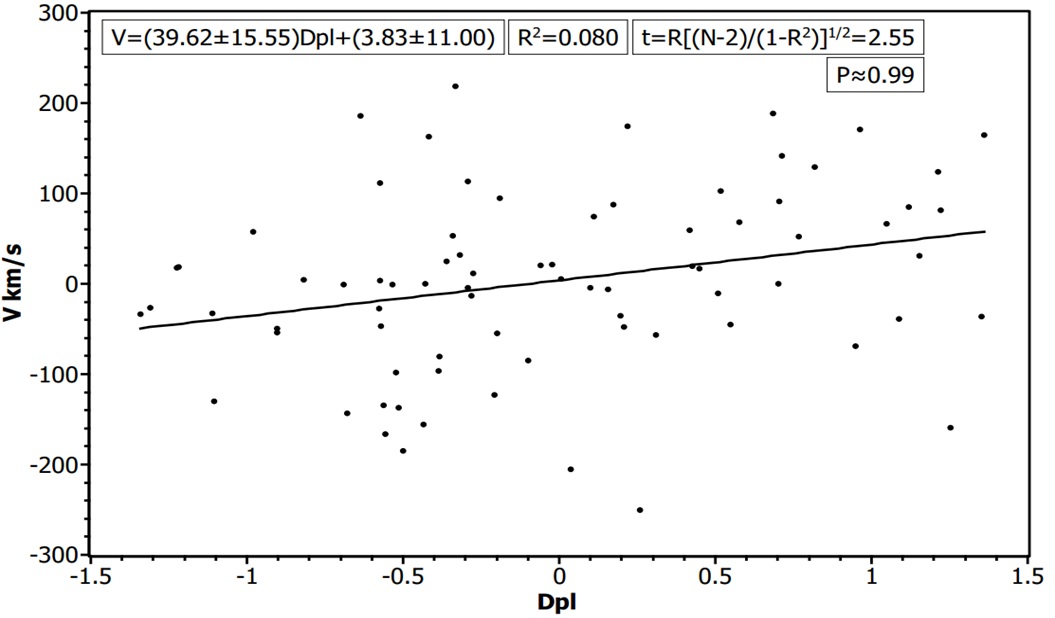}
    \caption{ Dependence of GC velocities in the direction perpendicular to the Galactic plane on the distance of the GC to the plane for the entire range -1.4 < Dpl < 1.4 kpc.}
    \label{fig:5}
\end{figure}
\newpage

\begin{longtable}{l c c c c c c r}    
\caption{The sample for 13°>b>-13° consists of 92 GCs} \label{tab:1} \\ \hline 
Name & RA & DEC & Rsun kpc & b deg & $\mu$l [mas yr-1] &  $\mu$b [mas yr-1] & V↑,V↓ km/s \\\hline
ESO\_452-SC11 & 249.854167 & -28.399167 & 7.39 & 12.10 & -5.8677450 & -3.0835660 & -102.18 \\
Rup\_106 & 189.667500 & -51.150278 & 20.71 & 11.67 & -1.2716863 & 0.3370817 & 24.67 \\
NGC\_6287 & 256.288904 & -22.708005 & 7.93 & 11.02 & -4.4497451 & 2.9439506 & 52.34 \\
NGC\_6333 & 259.799086 & -18.516257 & 8.3 & 10.71 & -3.8819845 & 0.0400754 & 59.27 \\
NGC\_5286 & 206.611710 & -51.374249 & 11.1 & 10.57 & 0.1481820 & -0.1964969 & 1.27 \\
NGC\_6535 & 270.960449 & -0.297639 & 6.36 & 10.44 & -4.5564066 & 2.3670137 & 31.31 \\
NGC\_6356 & 260.895804 & -17.813027 & 15.66 & 10.22 & -4.9125405 & 1.3039966 & 103.88 \\
NGC\_6284 & 256.120114 & -24.764799 & 14.21 & 9.94 & -3.5101230 & 1.4025257 & 98.06 \\
NGC\_6342 & 260.291573 & -19.587659 & 8.01 & 9.73 & -7.5343118 & -1.4921814 & -36.33 \\
NGC\_6273 & 255.657486 & -26.267971 & 8.34 & 9.38 & -0.6054084 & 3.6079414 & 164.55 \\
Ter\_3 & 247.162483 & -35.339829 & 7.64 & 9.20 & -5.0790522 & 2.8935819 & 81.81 \\
NGC\_3201 & 154.403427 & -46.412476 & 4.74 & 8.64 & 8.0250230 & 3.0374465 & 141.94 \\
Gran\_2 & 257.890000 & -24.849000 & 15.84 & 8.59 & -1.9641201 & -1.6698459 & -116.12 \\
NGC\_6779 & 289.148193 & 30.183472 & 10.43 & 8.34 & 0.5485667 & 2.5151089 & 103.26 \\
NGC\_6325 & 259.496327 & -23.767677 & 7.53 & 8.00 & -12.0979298 & 1.7598877 & 66.36 \\
NGC\_6293 & 257.542500 & -26.582083 & 9.19 & 7.83 & -3.0115925 & -3.2291540 & -159.04 \\
NGC\_6266 & 255.304153 & -30.113390 & 6.03 & 7.32 & -5.4375876 & 2.1811825 & 52.46 \\
NGC\_6139 & 246.918466 & -38.848782 & 10.04 & 6.94 & -6.1475598 & 2.5548822 & 123.74 \\
NGC\_6539 & 271.207276 & -7.585858 & 8.16 & 6.78 & -6.4126601 & 4.3415298 & 171.03 \\
NGC\_6517 & 270.460750 & -8.958778 & 9.23 & 6.76 & -4.6483533 & -0.7924843 & -38.58 \\
NGC\_6316 & 259.155417 & -28.140111 & 11.15 & 5.76 & -6.6254312 & 1.4319090 & 85.36 \\
IC\_1276 & 272.684441 & -7.207595 & 4.55 & 5.67 & -5.2491931 & 0.0734359 & 16.89 \\
NGC\_6355 & 260.993533 & -26.352827 & 8.65 & 5.43 & -3.1045387 & 3.6145579 & 129.12 \\
NGC\_6304 & 258.634399 & -29.462028 & 6.15 & 5.38 & -3.2467442 & 2.6895913 & 67.94 \\
NGC\_5927 & 232.002869 & -50.673031 & 8.27 & 4.86 & -5.9737147 & 0.2206639 & -0.19 \\
NGC\_5946 & 233.869051 & -50.659713 & 9.64 & 4.19 & -5.2698419 & 1.7848184 & 91.45 \\
NGC\_6401 & 264.652191 & -23.909605 & 7.44 & 3.98 & -0.2254790 & 3.1178677 & 102.45 \\
NGC\_6440 & 267.220167 & -20.360417 & 8.25 & 3.80 & -4.0480299 & -1.0291697 & -44.79 \\
Gran\_3 & 256.256000 & -35.496000 & 11.47 & 3.42 & -1.9279856 & 3.3706196 & 188.73 \\
NGC\_6256 & 254.886107 & -37.120968 & 7.24 & 3.31 & -3.5584548 & 1.8938272 & 59.18 \\
Pal\_10 & 289.508728 & 18.571667 & 8.94 & 2.72 & -8.3622493 & 0.4888675 & 19.20 \\
Ter\_2 & 261.887917 & -30.802333 & 7.75 & 2.30 & -6.3915840 & -1.6914029 & -56.78 \\
HP\_1 & 262.771667 & -29.981667 & 7 & 2.12 & -7.0349280 & -7.5988281 & -250.68 \\
Gran\_5 & 267.228000 & -24.170000 & 4.91 & 1.84 & -10.7390778 & -0.2004053 & -6.55 \\
Pal\_6 & 265.925812 & -26.224995 & 7.05 & 1.78 & -9.3142921 & 5.0606473 & 174.65 \\
Ter\_5 & 267.020200 & -24.779055 & 6.62 & 1.69 & -5.3372357 & -1.0392659 & -35.03 \\
Ter\_4 & 262.662506 & -31.595528 & 7.59 & 1.31 & -6.0171770 & 2.4625908 & 87.52 \\
Ter\_1 & 263.946667 & -30.481778 & 5.67 & 0.99 & -5.6574964 & -0.1965494 & -4.29 \\
VVV-CL001 & 268.677083 & -24.014722 & 8.08 & 0.78 & -3.1906301 & 2.0647555 & 74.68 \\
UKS\_1 & 268.613312 & -24.145277 & 15.58 & 0.76 & -3.0226820 & -0.6591470 & -47.92 \\
VVV-CL160 & 271.737500 & -20.011111 & 6.8 & 0.30 & -15.6280977 & -6.4001518 & -205.15 \\
2MASS-GC01 & 272.090851 & -19.829723 & 3.37 & 0.10 & -2.1965477 & 0.3430014 & 5.42 \\
Liller\_1 & 263.352333 & -33.389556 & 8.06 & -0.16 & -8.6757093 & 0.5495440 & 20.84 \\
2MASS-GC02 & 272.402100 & -20.778889 & 5.5 & -0.62 & -4.7191882 & 0.7329487 & 20.06 \\
FSR\_1716 & 242.625000 & -53.748889 & 7.43 & -1.59 & -9.2559262 & -3.5049283 & -122.63 \\
FSR\_1735 & 253.044174 & -47.058056 & 9.08 & -1.85 & -4.0720036 & 2.5688345 & 112.85 \\
Ter\_10 & 270.740833 & -26.066944 & 10.21 & -1.86 & -5.6075545 & 4.6487625 & 218.15 \\
Ter\_9 & 270.411667 & -26.839722 & 5.77 & -1.99 & -7.7646098 & -1.9157936 & -54.78 \\
Ter\_12 & 273.065833 & -22.741944 & 5.17 & -2.10 & -5.6765625 & 4.0026695 & 94.65 \\
Ter\_6 & 267.693250 & -31.275389 & 7.27 & -2.16 & -8.9621555 & 0.4812281 & 11.41 \\
NGC\_6544 & 271.833833 & -24.998222 & 2.58 & -2.20 & -17.3750724 & -7.0550301 & -84.80 \\
NGC\_6749 & 286.314056 & 1.899756 & 7.59 & -2.21 & -6.6322304 & -0.1851090 & -4.41 \\
Djor\_1 & 266.869583 & -33.066389 & 9.88 & -2.48 & -9.6932606 & -0.3274483 & 0.23 \\
Djor\_2 & 270.454378 & -27.825819 & 8.76 & -2.51 & -2.2577978 & -2.1068361 & -80.91 \\
Lynga\_7 & 242.765213 & -55.317776 & 7.9 & -2.80 & -7.6266323 & -2.5589874 & -96.65 \\
NGC\_6553 & 272.322992 & -25.908067 & 5.33 & -3.03 & -0.2035342 & -0.5159640 & -13.01 \\
FSR\_1758 & 262.800000 & -39.808000 & 11.09 & -3.29 & 0.5294556 & 3.7789394 & 185.42 \\
NGC\_6540 & 271.535657 & -27.765286 & 5.91 & -3.31 & -4.2579551 & 1.8611737 & 53.04 \\
NGC\_6380 & 263.618611 & -39.069530 & 9.61 & -3.42 & -3.9208846 & 0.0824420 & 3.84 \\
Ton\_2 & 264.042000 & -38.556100 & 6.99 & -3.42 & -3.8553524 & 4.5830534 & 162.72 \\
Patchick\_126 & 256.410833 & -47.342222 & 8 & -3.83 & -8.3804896 & -0.2312930 & -0.61 \\
NGC\_6453 & 267.715508 & -34.598477 & 10.07 & -3.87 & -5.0134890 & -3.1525475 & -143.55 \\
NGC\_6760 & 287.800268 & 1.030466 & 8.41 & -3.92 & -3.7289854 & -0.6933552 & -27.43 \\
NGC\_6522 & 270.891958 & -30.033974 & 7.29 & -3.93 & -4.3504848 & -5.3973421 & -185.16 \\
Gran\_1 & 269.651000 & -32.020000 & 8.27 & -3.98 & -10.9776707 & 2.9971787 & 111.82 \\
NGC\_6528 & 271.206697 & -30.055778 & 7.83 & -4.17 & -5.9878352 & -0.8433308 & -46.66 \\
BH\_140 & 193.472915 & -67.177276 & 4.81 & -4.31 & -14.8882993 & 1.4003266 & 25.08 \\
NGC\_6712 & 283.268021 & -8.705960 & 7.38 & -4.32 & -2.4494228 & -4.9969944 & -166.35 \\
NGC\_6838 & 298.443726 & 18.779194 & 4 & -4.56 & -4.0304540 & 1.5758897 & 31.61 \\
NGC\_6441 & 267.554413 & -37.051445 & 12.73 & -5.01 & -5.9319594 & -0.5151124 & -32.60 \\
BH\_261 & 273.527500 & -28.635000 & 6.12 & -5.27 & -1.4567879 & -4.8480501 & -134.64 \\
NGC\_6626 & 276.137039 & -24.869847 & 5.37 & -5.58 & -8.0707163 & -3.8364185 & -98.34 \\
NGC\_6558 & 272.573974 & -31.764508 & 7.79 & -6.02 & -4.4996159 & -0.4322626 & 4.59 \\
NGC\_6642 & 277.975957 & -23.475602 & 8.05 & -6.44 & -3.5592824 & -1.6037873 & -54.06 \\
NGC\_6569 & 273.411667 & -31.826889 & 10.53 & -6.68 & -8.4187435 & 0.2319725 & 17.31 \\
NGC\_6388 & 264.071777 & -44.735500 & 11.17 & -6.74 & -3.0005843 & -0.3197011 & -26.57 \\
Pal\_8 & 280.377290 & -19.828858 & 11.32 & -6.80 & -5.9339544 & -0.7175484 & -33.56 \\
NGC\_6638 & 277.733734 & -25.497473 & 9.78 & -7.15 & -4.7651555 & 0.4186629 & 18.20 \\
NGC\_6352 & 261.371277 & -48.422169 & 5.54 & -7.17 & -4.8921400 & -0.6296243 & -0.74 \\
NGC\_6656 & 279.099762 & -23.904749 & 3.3 & -7.55 & -0.6499270 & -11.3075781 & -155.93 \\
NGC\_6624 & 275.918793 & -30.361029 & 8.02 & -7.91 & -6.1470558 & -3.2581853 & -130.32 \\
NGC\_4833 & 194.891342 & -70.876503 & 6.48 & -8.02 & -8.4165855 & -0.7001526 & -49.48 \\
NGC\_4372 & 186.439101 & -72.659084 & 5.71 & -9.88 & -6.7104895 & 2.6446347 & 57.59 \\
NGC\_6496 & 269.765350 & -44.265945 & 9.64 & -10.01 & -9.6033491 & -1.6935433 & -52.84 \\
NGC\_6637 & 277.846252 & -32.348084 & 8.9 & -10.27 & -7.4403317 & 2.0260546 & 75.70 \\
NGC\_6717 & 283.775177 & -22.701473 & 7.52 & -10.90 & -5.8714507 & 0.7399592 & 20.20 \\
NGC\_6541 & 272.009827 & -43.714889 & 7.61 & -11.19 & -7.7409477 & -4.2616948 & -119.10 \\
NGC\_2808 & 138.012909 & -64.863495 & 10.06 & -11.25 & 0.4735823 & 0.9252551 & 23.09 \\
NGC\_6652 & 278.940125 & -32.990723 & 9.46 & -11.38 & -6.1790648 & 3.1761098 & 158.54 \\
NGC\_6397 & 265.175385 & -53.674335 & 2.48 & -11.96 & -13.6757610 & -11.6202306 & -137.57 \\
NGC\_6681 & 280.803162 & -32.292110 & 9.36 & -12.51 & -3.7237616 & -3.2439020 & -187.53 \\
ESO\_280-SC06 & 272.275000 & -46.423333 & 20.95 & -12.57 & -2.7963703 & -0.6078765 & -79.25 \\\hline

\end{longtable}

\begin{longtable}{l c c c c c c r}    
\caption{The GC sample from the general table for the range of distances to the plane from 0 to 1.4 kpc contains 77 GCs} \label{tab:2} \\ \hline   
Name & RA & DEC & Rsun kpc & <RV> km/s &  $\mu$b [mas yr-1] & Dpl kpc & V↑,V↓ km/s \\\hline 
NGC\_6273 & 255.657486 & -26.267971 & 8.34 & 145.54 & 3.6079414229 & 1.360 & 165 \\
NGC\_6342 & 260.291573 & -19.587659 & 8.01 & 115.75 & -1.4921814347 & 1.353 & -36 \\
NGC\_6293 & 257.542500 & -26.582083 & 9.19 & -143.66 & -3.2291539667 & 1.253 & -159 \\
Ter\_3 & 247.162483 & -35.339829 & 7.64 & -135.76 & 2.8935818900 & 1.221 & 82 \\
NGC\_6139 & 246.918466 & -38.848782 & 10.04 & 24.41 & 2.5548821981 & 1.213 & 124 \\
NGC\_6535 & 270.960449 & -0.297639 & 6.36 & -214.85 & 2.3670136697 & 1.152 & 31 \\
NGC\_6316 & 259.155417 & -28.140111 & 11.15 & 99.65 & 1.4319090063 & 1.120 & 85 \\
NGC\_6517 & 270.460750 & -8.958778 & 9.23 & -35.06 & -0.7924842869 & 1.087 & -39 \\
NGC\_6325 & 259.496327 & -23.767677 & 7.53 & 29.54 & 1.7598877255 & 1.048 & 66 \\
NGC\_6539 & 271.207276 & -7.585858 & 8.16 & 35.19 & 4.3415298028 & 0.963 & 171 \\
NGC\_6366 & 261.934357 & -5.079861 & 3.44 & -120.65 & -2.2913000101 & 0.950 & -69 \\
NGC\_6355 & 260.993533 & -26.352827 & 8.65 & -195.85 & 3.6145578895 & 0.818 & 129 \\
NGC\_6266 & 255.304153 & -30.113390 & 6.03 & -73.98 & 2.1811824878 & 0.768 & 52 \\
NGC\_3201 & 154.403427 & -46.412476 & 4.74 & 495.38 & 3.0374464690 & 0.712 & 142 \\
NGC\_5946 & 233.869051 & -50.659713 & 9.64 & 137.6 & 1.7848183790 & 0.704 & 91 \\
NGC\_5927 & 232.002869 & -50.673031 & 8.27 & -104.09 & 0.2206639186 & 0.701 & 0 \\
Gran\_3 & 256.256000 & -35.496000 & 11.47 & 94.87 & 3.3706196353 & 0.685 & 189 \\
NGC\_6304 & 258.634399 & -29.462028 & 6.15 & -108.62 & 2.6895912823 & 0.576 & 68 \\
NGC\_6440 & 267.220167 & -20.360417 & 8.25 & -69.39 & -1.0291696623 & 0.547 & -45 \\
NGC\_6401 & 264.652191 & -23.909605 & 7.44 & -105.44 & 3.1178677174 & 0.516 & 102 \\
NGC\_6121 & 245.896744 & -26.525749 & 1.85 & 71.22 & -3.5435439040 & 0.509 & -10 \\
IC\_1276 & 272.684441 & -7.207595 & 4.55 & 155.06 & 0.0734358716 & 0.449 & 17 \\
Pal\_10 & 289.508728 & 18.571667 & 8.94 & -31.7 & 0.4888674927 & 0.425 & 19 \\
NGC\_6256 & 254.886107 & -37.120968 & 7.24 & -99.75 & 1.8938272225 & 0.418 & 59 \\
Ter\_2 & 261.887917 & -30.802333 & 7.75 & 133.46 & -1.6914029127 & 0.311 & -57 \\
HP\_1 & 262.771667 & -29.981667 & 7 & 39.76 & -7.5988280550 & 0.258 & -251 \\
Pal\_6 & 265.925812 & -26.224995 & 7.05 & 177 & 5.0606473239 & 0.219 & 175 \\
UKS\_1 & 268.613312 & -24.145277 & 15.58 & 59.38 & -0.6591469714 & 0.208 & -48 \\
Ter\_5 & 267.020200 & -24.779055 & 6.62 & -81.97 & -1.0392659446 & 0.195 & -35 \\
Ter\_4 & 262.662506 & -31.595528 & 7.59 & -48.96 & 2.4625907937 & 0.173 & 88 \\
Gran\_5 & 267.228000 & -24.170000 & 4.91 & -58.87 & -0.2004052860 & 0.158 & -7 \\
VVV-CL001 & 268.677083 & -24.014722 & 8.08 & -327.28 & 2.0647554734 & 0.110 & 75 \\
Ter\_1 & 263.946667 & -30.481778 & 5.67 & 57.55 & -0.1965494344 & 0.098 & -4 \\
VVV-CL160 & 271.737500 & -20.011111 & 6.8 & 245.28 & -6.4001518435 & 0.036 & -205 \\
2MASS-GC01 & 272.090851 & -19.829723 & 3.37 & -35.58 & 0.3430013522 & 0.006 & 5 \\
Liller\_1 & 263.352333 & -33.389556 & 8.06 & 60.36 & 0.5495440375 & -0.023 & 21 \\
2MASS-GC02 & 272.402100 & -20.778889 & 5.5 & -87.5 & 0.7329486715 & -0.059 & 20 \\
NGC\_6544 & 271.833833 & -24.998222 & 2.58 & -38.46 & -7.0550300926 & -0.099 & -85 \\
Ter\_12 & 273.065833 & -22.741944 & 5.17 & 94.01 & 4.0026694574 & -0.190 & 95 \\
Ter\_9 & 270.411667 & -26.839722 & 5.77 & 68.56 & -1.9157935705 & -0.200 & -55 \\
FSR\_1716 & 242.625000 & -53.748889 & 7.43 & -30.7 & -3.5049282823 & -0.206 & -123 \\
Ter\_6 & 267.693250 & -31.275389 & 7.27 & 137.15 & 0.4812280920 & -0.274 & 11 \\
NGC\_6553 & 272.322992 & -25.908067 & 5.33 & -0.27 & -0.5159639546 & -0.282 & -13 \\
NGC\_6749 & 286.314056 & 1.899756 & 7.59 & -58.44 & -0.1851090091 & -0.292 & -4 \\
FSR\_1735 & 253.044174 & -47.058056 & 9.08 & -69.85 & 2.5688344515 & -0.294 & 113 \\
NGC\_6838 & 298.443726 & 18.779194 & 4 & -22.72 & 1.5758897311 & -0.318 & 32 \\
Ter\_10 & 270.740833 & -26.066944 & 10.21 & 211.37 & 4.6487625204 & -0.332 & 218 \\
NGC\_6540 & 271.535657 & -27.765286 & 5.91 & -16.5 & 1.8611736763 & -0.342 & 53 \\
BH\_140 & 193.472915 & -67.177276 & 4.81 & 90.3 & 1.4003266259 & -0.361 & 25 \\
Djor\_2 & 270.454378 & -27.825819 & 8.76 & -149.75 & -2.1068361326 & -0.383 & -81 \\
Lynga\_7 & 242.765213 & -55.317776 & 7.9 & 17.86 & -2.5589874141 & -0.386 & -97 \\
Ton\_2 & 264.042000 & -38.556100 & 6.99 & -184.72 & 4.5830533729 & -0.417 & 163 \\
Djor\_1 & 266.869583 & -33.066389 & 9.88 & -359.18 & -0.3274483325 & -0.428 & 0 \\
NGC\_6656 & 279.099762 & -23.904749 & 3.3 & -148.72 & -11.3075780998 & -0.434 & -156 \\
NGC\_6522 & 270.891958 & -30.033974 & 7.29 & -15.23 & -5.3973420677 & -0.499 & -185 \\
NGC\_6397 & 265.175385 & -53.674335 & 2.48 & 18.51 & -11.6202306433 & -0.514 & -138 \\
NGC\_6626 & 276.137039 & -24.869847 & 5.37 & 11.11 & -3.8364185245 & -0.522 & -98 \\
Patchick\_126 & 256.410833 & -47.342222 & 8 & -122.14 & -0.2312929821 & -0.534 & -1 \\
NGC\_6712 & 283.268021 & -8.705960 & 7.38 & -107.45 & -4.9969944420 & -0.556 & -166 \\
BH\_261 & 273.527500 & -28.635000 & 6.12 & -60 & -4.8480501092 & -0.562 & -135 \\
NGC\_6528 & 271.206697 & -30.055778 & 7.83 & 211.86 & -0.8433307918 & -0.570 & -47 \\
Gran\_1 & 269.651000 & -32.020000 & 8.27 & 78.94 & 2.9971787050 & -0.574 & 112 \\
NGC\_6380 & 263.618611 & -39.069530 & 9.61 & -1.48 & 0.0824420498 & -0.574 & 4 \\
NGC\_6760 & 287.800268 & 1.030466 & 8.41 & -2.37 & -0.6933551606 & -0.576 & -27 \\
FSR\_1758 & 262.800000 & -39.808000 & 11.09 & 227.31 & 3.7789394057 & -0.637 & 185 \\
NGC\_6453 & 267.715508 & -34.598477 & 10.07 & -99.23 & -3.1525475223 & -0.680 & -144 \\
NGC\_6352 & 261.371277 & -48.422169 & 5.54 & -125.63 & -0.6296242777 & -0.691 & -1 \\
NGC\_6558 & 272.573974 & -31.764508 & 7.79 & -195.12 & -0.4322626027 & -0.817 & 5 \\
NGC\_6642 & 277.975957 & -23.475602 & 8.05 & -60.61 & -1.6037873078 & -0.903 & -54 \\
NGC\_4833 & 194.891342 & -70.876503 & 6.48 & 201.99 & -0.7001526401 & -0.904 & -49 \\
NGC\_4372 & 186.439101 & -72.659084 & 5.71 & 75.59 & 2.6446347474 & -0.980 & 58 \\
NGC\_6624 & 275.918793 & -30.361029 & 8.02 & 54.79 & -3.2581853134 & -1.104 & -130 \\
NGC\_6441 & 267.554413 & -37.051445 & 12.73 & 18.47 & -0.5151123783 & -1.111 & -33 \\
NGC\_6638 & 277.733734 & -25.497473 & 9.78 & 8.63 & 0.4186628947 & -1.218 & 18 \\
NGC\_6569 & 273.411667 & -31.826889 & 10.53 & -49.83 & 0.2319725137 & -1.225 & 17 \\
NGC\_6388 & 264.071777 & -44.735500 & 11.17 & 83.11 & -0.3197010757 & -1.311 & -27 \\
Pal\_8 & 280.377290 & -19.828858 & 11.32 & -39.68 & -0.7175484457 & -1.340 & -34 \\\hline

\end{longtable}

\end{document}